\documentclass[preprint,showpacs,preprintnumbers,amsmath,amssymb]{revtex4}

\usepackage{graphicx}% Include figure files
\usepackage{dcolumn}% Align table columns on decimal point
\usepackage{bm}% bold math

\begin{document}

\preprint{Submission to Phys. Rev. B}

\title{
Experimental confirmation of spin gap in 
antiferromagnetic alternating spin-3/2 chain 
substances {\it R}CrGeO$_5$ ({\it R} = Y or $^{154}$Sm) 
by inelastic neutron scattering experiments
}

\author{Masashi Hase$^1$}
 \email{HASE.Masashi@nims.go.jp}
\author{Minoru Soda$^2$}
\author{Takatsugu Masuda$^2$}
\author{Daichi Kawana$^{2,3}$}
\author{Tetsuya Yokoo$^3$}
\author{Shinichi Itoh$^3$}
\author{Akira Matsuo$^2$}
\author{Koichi Kindo$^2$}
\author{Masanori Kohno$^1$}

\affiliation{%
${}^{1}$National Institute for Materials Science (NIMS), 
Tsukuba, Ibaraki 305-0047, Japan \\
${}^{2}$ The Institute for Solid State Physics (ISSP), 
the University of Tokyo, 
Kashiwa, Chiba 277-8581, Japan \\
${}^{3}$High Energy Accelerator Research Organization (KEK), 
Tsukuba, Ibaraki 305-0801, Japan 
}%

\date{\today}% It is always \today, today,

\begin{abstract}

A spin-singlet ground state with a spin gap 
has been discovered in 
antiferromagnetic spin chain substances 
when the spin value is 1/2, 1 or 2.
To find spin gap (singlet-triplet) excitations in 
spin-3/2 chain substances, 
we performed inelastic neutron scattering and 
magnetization measurements 
on {\it R}CrGeO$_5$ ({\it R} = Y or Sm) powders. 
As expected, 
we observed spin gap excitations and  
the dispersion relation of the lowest magnetic excitations.  
We proved that the spin system of Cr$^{3+}$ was 
an antiferromagnetic alternating spin-3/2 chain. 

\end{abstract}


\maketitle

\section{INTRODUCTION}

The ground state (GS) is spin singlet in 
antiferromagnetic (AF) Heisenberg alternating spin chains 
because of large quantum fluctuation. 
The Hamiltonian is given as follows. 
\begin{equation}
{\cal H} = J \sum_{i} [1-(-1)^i \delta] {\bm S}_{i} \cdot {\bm S}_{i+1}.
\end{equation}
In the AF Heisenberg uniform ($\delta = 0$) spin-1/2 chain, 
GS is almost an ordered state in spite of spin singlet (critical state)  
and is designated the gapless Tomonaga-Luttinger liquid (TLL). 
A gapless GS can also appear when the spin value $S$ is larger than 
1/2.\cite{Kato94,Kohno98,Yajima96,Yamamoto97} 
The gapless GS is regarded as TLL.  
A spin gap (singlet-triplet gap) opens except for gapless point(s) 
and a spin-singlet GS is stabilized. 
The spin-singlet GS can be expressed 
using valence-bond-solid (VBS) diagrams, as depicted in 
Fig. 1.\cite{Yajima96} 
Considering the range of $\delta$ in which VBS exists, 
the appearance of the spin gap is a common phenomenon 
in AF alternating spin chains. 

\begin{figure}
\begin{center}
\includegraphics[width=8cm]{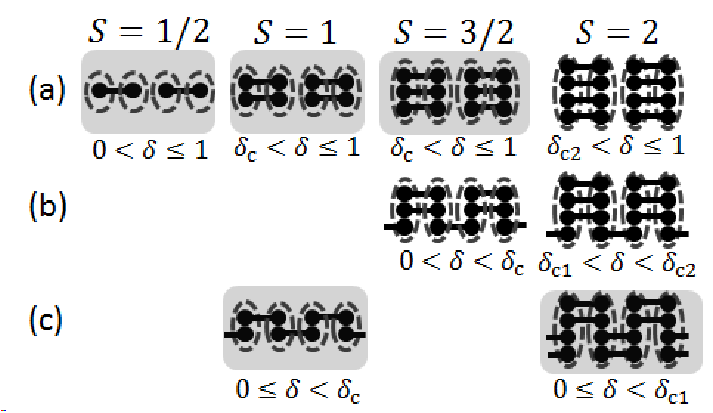}
\caption{
(Color online)
The valence-bond-solid (VBS) diagrams.\cite{Yajima96} 
The spin-singlet ground state shown in (a), (b), and (c) is designated 
$(2S, 0)$, $(2S-1, 1)$, and $(S, S)$ states, respectively.
Small circles and lines represent 
spin-1/2 variables and singlet pairs, respectively. 
Each ellipse represents the symmetrization of the spin-1/2 variables 
on each site to create the total spin variable.
The ground states are accompanied with a spin gap. 
The value of $\delta$ of gapless points is 
0 for $S=1/2$, 
$\delta_{\rm c} = 0.2595(5)$ for $S=1$,\cite{Kohno98} 
0 and $\delta_{\rm c} = 0.42(2)$ for $S=3/2$,\cite{Yajima96} and 
$\delta_{\rm c1} = 0.18(1)$ and 
$\delta_{\rm c2} = 0.545(5)$ for $S=2$.\cite{Yamamoto97}
Model substances were found for the hatched GSs 
when $S=1/2, 1$, and 2, as described in the text. 
In this study, we show that 
{\it R}CrGeO$_5$ ({\it R} = Y or Sm) are model substances 
for the hatched GS of $S=3/2$ (a). 
}
\end{center}
\end{figure}

Model substances have been found for the hatched GSs  in Fig. 1
when $S=1/2, 1$, and 2. 
Cu(NO$_3$)$_2$-2.5H$_2$O,\cite{Diederix79,Bonner83} 
TTF-{\it M}S$_4$C$_4$(CF$_3$)$_4$ 
({\it M}= Cu or Au, TTF=tetrathiafulvalene),\cite{Bray75,Jacobs76}
and CuGeO$_3$\cite{Hase93a,Hase93b,Hase93c} 
are model substances for $S = 1/2$ (a). 
[Ni({\it N, N'}-bis(3-aminopropyl)propane-1, 3-diamine($\mu$-NO$_2$)]ClO$_4$ 
(abbreviated as NTENP).\cite{Narumi01,Zheludev04} 
is a model substance for $S = 1$ (a).  
Ni(C$_2$H$_8$N$_2$)$_2$NO$_2$(ClO$_4$) 
(abbreviated as NENP)\cite{Renard87} 
and Y$_2$BaNiO$_5$\cite{Darriet93,DiTusa94,Yokoo95}
are model substances for $S = 1$ (c). 

When the spin value is larger than 1, 
almost no model substance exists. 
The only example reported in the literature 
is MnCl$_3$(C$_{10}$H$_8$N$_2$).\cite{Granroth96}
This substance has an AF uniform spin-2 chain of which 
GS is shown by the diagram of $S = 2$ (c). 
The energy gap was evaluated as 0.32(8) meV and 0.14(3) meV 
from magnetization curves at 30 mK 
in the magnetic field parallel and perpendicular to chains, respectively.  
These gaps are consistent with a Haldane gap of 0.20(7) meV, 
where the excited triplet is split by single-ion anisotropy $D = 0.03(1)$ meV.
The temperature $T$ dependence of magnetic susceptibility and 
the magnetization curve were well fitted to calculated results 
with $J=31.2$ K and the $g$-value of 2.02.\cite{Hagiwara12}
Inelastic neutron scattering results 
provide microscopic evidence for the presence of the Haldane gap.\cite{Granroth02}

We comment on AF Heisenberg uniform spin chain substances
{\it AMX}$_3$. 
Here, {\it A} is K, Rb or Cs, {\it M} is 3{\it d} atom, and 
{\it X} is F, Cl or Br.
In the uniform spin-1 chain substance CsNiCl$_3$, 
the Haldane gap was observed in inelastic neutron scattering 
experiments.\cite{Kakurai91}
In the uniform spin-3/2 chain substances CsVCl$_3$ and CsVBr$_3$, 
low-energy broad excitations were observed at the magnetic zone center 
at 20 K ($> T_{\rm N} = 13.3$ K)\cite{Itoh12a} 
and 25 K ($> T_{\rm N} = 20.3$ K),\cite{Itoh99} respectively, 
where $T_{\rm N}$ is an AF transition temperature. 
The low-energy broad excitations were recognized as a part of the continuum 
just above the lowest magnetic excitations. 
Therefore, the experimental results do not contradict  
the theoretical prediction (gapless excitations).
In the uniform spin-2 chain substance CsCrCl$_3$, 
no excitation gap was detected at the magnetic zone center 
at 20 K ($> T_{\rm N} = 16$ K)
within the experimental errors.\cite{Itoh02} 
The Haldane gap can be estimated to be 0.2 meV and 
is predicted to be observable below 1.4 K. 
Thus, the Haldane gap were unable to be detected in the experiments. 

We can show experimentally that the appearance of the spin gap 
is a universal phenomenon irrespective of the spin value below 2
if we can find the spin gap in an AF alternating spin-3/2 chain substance. 
We have devoted attention to 
insulating {\it R}CrGeO$_5$ ({\it R} = Y or rare earth) 
as shown in Fig. 2.\cite{Shpanchenko08} 
A Cr$^{3+}$ ion is surrounded by O$^{2-}$ ligands and 
forms a CrO$_6$ octahedron. 
In the ground state of Cr$^{3+}$ ions, 
the orbital degree of freedom is quenched. 
Spin 3/2 is responsible for the magnetism of Cr$^{3+}$ ions. 
From the crystal structure, 
Cr$^{3+}$ spins are expected to form an alternating spin-3/2 chain.
Table I shows the Cr-O-Cr angle and Cr-Cr distance in 
two kinds of Cr-Cr bonds ($d_1$ and $d_2$ bonds).
Spin chains are separated from one another 
by GeO$_5$ square pyramids and R$^{3+}$ ions.
Shpanchenko {\it et al.} reported the $T$ dependence of 
magnetic susceptibility ($\chi$) of 
{\it R}CrGeO$_5$ ({\it R} = Sm, Eu or Nd) powders.\cite{Shpanchenko08} 
A broad maximum of $\chi$ appears at $T_{\rm max} = 220$ K and 100 K 
in SmCrGeO$_5$ and EuCrGeO$_5$, respectively,  
indicating the existence of a low-dimensional AF spin system.   
Considering the crystal structure and the large values of $T_{\rm max}$, 
we can expect that 
Cr$^{3+}$ spins form an AF Heisenberg alternating spin-3/2 chain.
No magnetic transition appears down to 1.8 K in the two substances.
In NdCrGeO$_5$, $\chi$ of Nd$^{3+}$ ions is very large.
We cannot determine 
whether $\chi$ of Cr$^{3+}$ spins shows a broad maximum or not. 
Probably, an AF alternating spin-3/2 chain also exists in NdCrGeO$_5$ 
because of the same crystal structure. 
A clear peak appears at 2.6 K in $\chi$ of NdCrGeO$_5$. 
The susceptibility of Nd$^{3+}$ ions is dominant at low $T$. 
Probably, Nd$^{3+}$ magnetic moments generate a magnetic transition.

\begin{figure}
\begin{center}
\includegraphics[width=8cm]{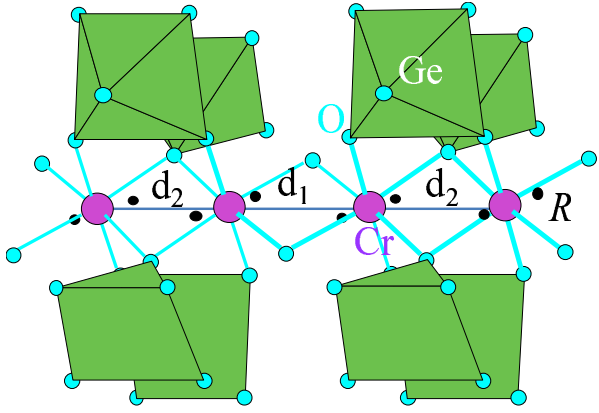}
\caption{
(Color online)
A part of the {\it R}CrGeO$_5$ structure showing 
two kinds of Cr-Cr bonds in the chain of edge-sharing CrO$_6$ octahedra,  
GeO$_5$ square pyramids, and {\it R} atoms. 
}
\end{center}
\end{figure}

As $T$ is lowered, the susceptibility of {\it R}CrGeO$_5$ ({\it R} = Sm, Eu or Nd)
increases like a Curie-Weiss susceptibility. 
The increase originates in rare earth ions or magnetic other materials. 
The susceptibility becomes nearly zero at low $T$ 
if a spin-singlet ground state with a spin gap exists in the Cr$^{3+}$ spin system. 
Because of the Curie-Weiss susceptibility at low $T$, unfortunately, 
we cannot prove a spin-singlet ground state with a spin gap from the susceptibility results.
Consequently, we performed 
inelastic neutron scattering (INS) and magnetization measurements 
on {\it R}CrGeO$_5$ ({\it R} = Y or Sm) powders 
to confirm the spin gap (singlet-triplet) excitations. 

\begin{table*}
\caption{\label{table1}
Cr-O-Cr angles and Cr-Cr distances in two kinds of Cr-Cr bonds 
($d_1$ and $d_2$ bonds) in 
{\it R}CrGeO$_5$ ({\it R} = Y or Sm).\cite{Shpanchenko08}
}
\begin{ruledtabular}
\begin{tabular}{cccc}
& & Y & Sm \\
\hline
$d_1$ bond & Cr-O-Cr angle & 95.9$^{\circ}$ & 97.3$^{\circ}$ \\
& Cr-Cr distance & 2.872 \AA & 2.952 \AA \\
\hline
$d_2$ bond & Cr-O-Cr angle & 92.8$^{\circ}$ & 91.3$^{\circ}$ \\
& Cr-Cr  distance & 2.811 \AA & 2.770 \AA \\
\end{tabular}
\end{ruledtabular}
\end{table*}

\section{Methods of Experiments and Calculations}

Crystalline powders of {\it R}CrGeO$_5$ ({\it R} = Y or Sm)
were synthesized using a solid-state-reaction method 
at 1,523 K in air with intermediate grindings.\cite{Shpanchenko08} 
We used an isotope $^{154}$Sm (purity of the isotope: 99 \%) 
for powders of INS experiments 
to decrease absorption of neutrons. 
We confirmed formation of {\it R}CrGeO$_5$ ({\it R} = Y or Sm) 
using an X-ray diffractometer (RINT-TTR III; Rigaku).
We were able to obtain samples of a nearly single phase of SmCrGeO$_5$.
We found the existence of non-magnetic Y$_2$Ge$_2$O$_7$ 
in diffraction patterns of YCrGeO$_5$ samples. 
The molar ratio of Y$_2$Ge$_2$O$_7$ 
was estimated roughly as 10 \% \ from diffraction intensities. 

We measured magnetizations up to $H = 5$ T 
using a superconducting quantum interference device 
(SQUID) magnetometer (MPMS-5S; Quantum Design). 
High-field magnetization measurements were conducted 
using an induction method with a multilayer pulsed field magnet
installed at the Institute for Solid State Physics (ISSP), 
the University of Tokyo. 
We performed inelastic neutron scattering measurements 
using the High Resolution Chopper (HRC) spectrometer at BL 12 
in Japan Proton Accelerator Research Complex 
(J-PARC).\cite{Itoh11,Yano11,Itoh12b}
The energy resolution at 
the energy transfer $\omega = 0$ meV is 
3 - 5 \% of $E_{\rm i}$ (the energy of incident neutrons).
The $Q$ resolution is better than 0.1 \AA$^{-1}$ %$
where $Q$ is the magnitude of the scattering vector.

We calculated susceptibility of AF Heisenberg  alternating spin-3/2 chains  [Eq. (1)] 
using the quantum Monte Carlo loop algorithm\cite{Evertz03}
on 240 site chains. 
Finite-size effects and statistical errors are negligible 
in the scales of figures represented in this paper. 
We calculated the dynamical structure factor, which is proportional to neutron scattering intensity, on 120 site chains 
under the open boundary condition using the dynamical density-matrix renormalization group (DMRG) 
method.\cite{Jeckelmann02}

\section{Results and discussion}

We show $\chi$ of YCrGeO$_5$ and SmCrGeO$_5$ powders as red circles 
in Figs. 3(a) and 3(b), respectively. 
The value of the applied magnetic field is 0.01 T. 
The molar ratio of Y$_2$Ge$_2$O$_7$ included in the YCrGeO$_5$ sample
was estimated as 8.8 \%. 
Considering the Y$_2$Ge$_2$O$_7$ weight, 
we obtained the susceptibility in Fig. 3(a).
The susceptibility of YCrGeO$_5$ at low $T$ 
increases like a Curie-Weiss susceptibility 
probably because of unidentified magnetic material(s) in the YCrGeO$_5$ sample. 
We were unable to prove a spin-singlet ground state with a spin gap 
because of the Curie-Weiss susceptibility. 
No magnetic transition appears down to 2 K.
The susceptibility of our SmCrGeO$_5$ powders agrees with 
that reported by Shpanchenko {\it et al.}\cite{Shpanchenko08}
Results of analyses are described later.

\begin{figure}
\begin{center}
\includegraphics[width=8cm]{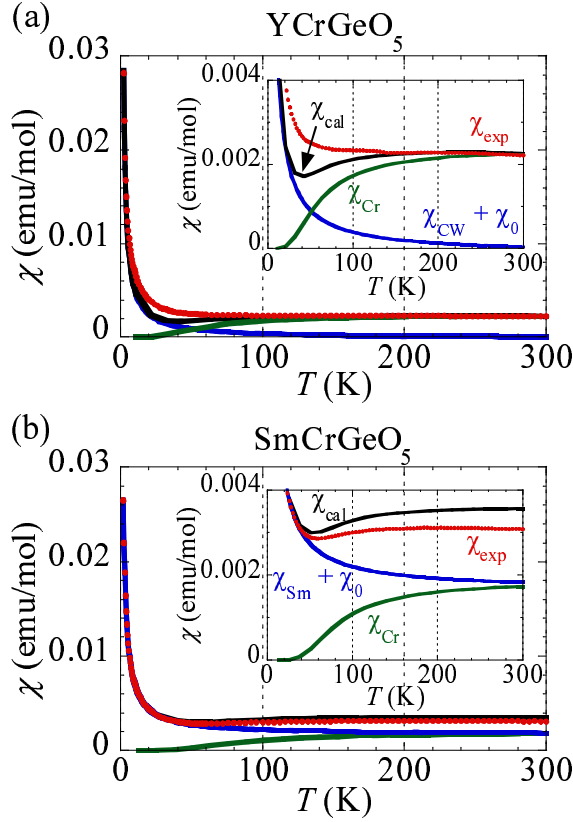}
\caption{
(Color online)
Magnetic susceptibility in 0.01 T of YCrGeO$_5$ (a) and SmCrGeO$_5$ (b).
The insets represent $\chi$ below 0.004 emu/mol. 
Circles show the experimental results. 
Three lines in each figure are explained in the text.
}
\end{center}
\end{figure}

We show high-field magnetizations at 4.2 K
of YCrGeO$_5$ and SmCrGeO$_5$ powders  
in Figs. 4(a) and 4(b), respectively. 
An upturn of the magnetization of YCrGeO$_5$ 
is apparent around 55 T, 
suggesting spin-gap closing induced by the magnetic field. 
The spin-gap closing might occur in higher fields. 
The slope of the magnetization is small around 
$M = 0.15$ $\mu_{\rm B}$/formula unit, 
indicating that about 5 \% Cr$^{3+}$ spins are nearly isolated.
We did not observe an upturn in the magnetization of SmCrGeO$_5$
up to 58 T.
The magnetization results suggest that 
YCrGeO$_5$ has a smaller spin gap than SmCrGeO$_5$. 
This point is consistent with INS results presented below.

\begin{figure}
\begin{center}
\includegraphics[width=8cm]{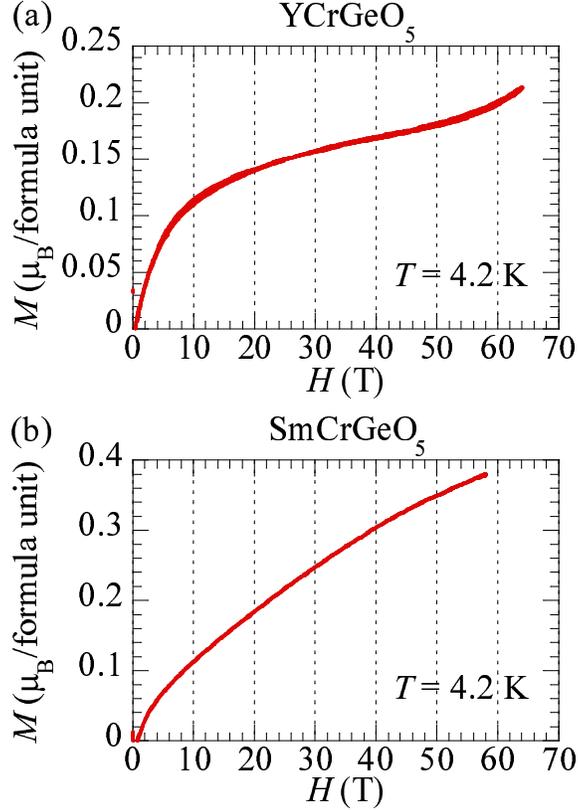}
\caption{
(Color online)
High-field magnetization at 4.2 K of YCrGeO$_5$ (a) and 
SmCrGeO$_5$ (b).
}
\end{center}
\end{figure}

We show INS results of YCrGeO$_5$ at 4.0 K and 199 K  
in Figs. 5(a) and 5(b), respectively, and 
those of $^{154}$SmCrGeO$_5$ at 7.8 K and 202 K  
in Figs. 6(a) and 6(b), respectively.
The energies of incident neutrons $E_{\rm i}$ are 51.1 and 91.6 meV 
for the measurements of YCrGeO$_5$ and $^{154}$SmCrGeO$_5$, respectively.
In YCrGeO$_5$, excitations are observed in the energy range of
8 meV $\lesssim \omega \lesssim 23$ meV
at 4.0 K and the intensity decreases with the increase of $Q$.
The intensity of the excitations is suppressed at higher temperature, 199 K.
The results mean that 
the observed excitations are dominated by those of magnetic origin.
Furthermore no excitation is observed at $\omega \lesssim$ 8 meV  and
this means the existence of spin gap.
Qualitatively the same behaviors are observed in $^{154}$SmCrGeO$_5$.
YCrGeO$_5$ and $^{154}$SmCrGeO$_5$ are
the first 
spin-3/2 chain substances having a spin gap.

\begin{figure}
\begin{center}
\includegraphics[width=8cm]{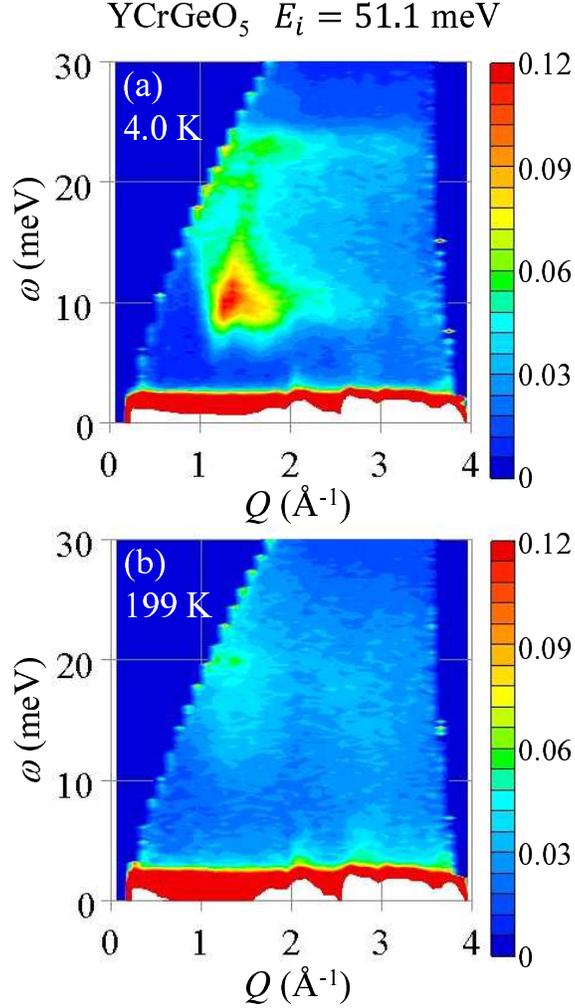}
\caption{
(Color online)
Maps of neutron scattering intensity in the $Q - \omega$ plane 
of YCrGeO$_5$ at 4.0 K (a) and 199 K (b). 
The energy of incident neutrons $E_{\rm i}$ is 51.1 meV. 
The numbers of protons injected to the neutron production target
are about $1.47 \times 10^{19}$ and $1.30 \times 10^{19}$ 
for the measurements at 4.0 K and 199 K, respectively. 
When the beam power is 200 kW, 
the total number of protons per day is $3.6 \times 10^{19}$. 
The spent times are about 9.8 and 8.7 hr 
for the measurements at 4.0 K and 199 K, respectively. 
The right vertical keys show the INS intensity in arbitrary units. 
The intensity is normalized 
to compare two data in different proton numbers. 
}
\end{center}
\end{figure}

\begin{figure}
\begin{center}
\includegraphics[width=8cm]{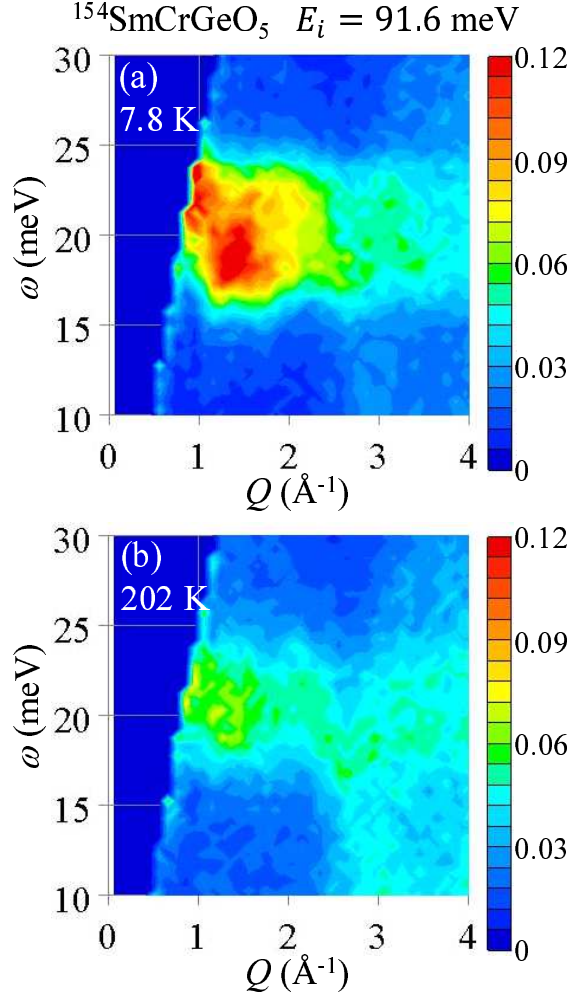}
\caption{
(Color online)
Maps of neutron scattering intensity in the $Q - \omega$ plane 
of $^{154}$SmCrGeO$_5$ at 7.8 K (a) and 202 K (b). 
The energy of incident neutrons $E_{\rm i}$ is 91.6 meV. 
The numbers of protons injected to the neutron production target
are about $2.79 \times 10^{19}$ and $2.60 \times 10^{19}$ 
for the measurements at 7.8 K and 202 K, respectively. 
The spent times are about 18.6 and 17.3 hr 
for the measurements at 7.8 K and 202 K, respectively. 
The right vertical keys show the INS intensity in arbitrary units. 
The intensity is normalized 
to compare two data in different proton numbers. 
}
\end{center}
\end{figure}

We obtained intensity maps in the $k - \omega$ plane 
as shown in Figs. 7(a) and 8(a) 
from the low $T$ data 
using the conversion method developed by Tomiyasu {\it et al}.\cite{Tomiyasu09}
The formula is given as follows.
\begin{equation}
S_{\rm SX}^{\rm (1D)}(Q_{\rm 1D}, \omega)=
[ S_{\rm pwd}(Q, \omega) + 
Q \frac{\partial S_{\rm pwd}(Q, \omega)}{\partial Q}
]_{Q=Q_{\rm 1D}}.
\end{equation}
Here, $Q_{\rm 1D}$, $S_{\rm pwd}(Q, \omega)$, and 
$S_{\rm SX}^{\rm (1D)}(Q_{\rm 1D}, \omega)$ represent 
the magnitude of the scattering vector parallel to the spin chain, 
the powder average scattering function, and 
the scattering function for $Q_{\rm 1D}$ expected in a single crystal, 
respectively. 
The normalized wavenumber $k$ is defined as 
$Q_{\rm 1D} \frac{d_1 + d_2}{2}$. 
The values of $d_1 + d_2$ at room temperature 
are 5.68 and 5.72 \AA \ for YCrGeO$_5$ and SmCrGeO$_5$, 
respectively.\cite{Shpanchenko08}
The intensity is the strongest at around $k = \pi$,
as expected in AF alternating spin chains. 
The magnetic excitations seem to have a dispersion relation. 
The white line in Figs. 7(a) and 8(a) shows 
the empirical dispersion relation of the lowest magnetic excitations 
$\omega (k) = \sqrt{A^2 \sin^2 k + \Delta^2}$. 
The values are 
$\Delta = 10$ meV and $A = 20$ meV in YCrGeO$_5$, and 
$\Delta = 18$ meV and $A = 15$ meV in $^{154}$SmCrGeO$_5$.

\begin{figure}
\begin{center}
\includegraphics[width=8cm]{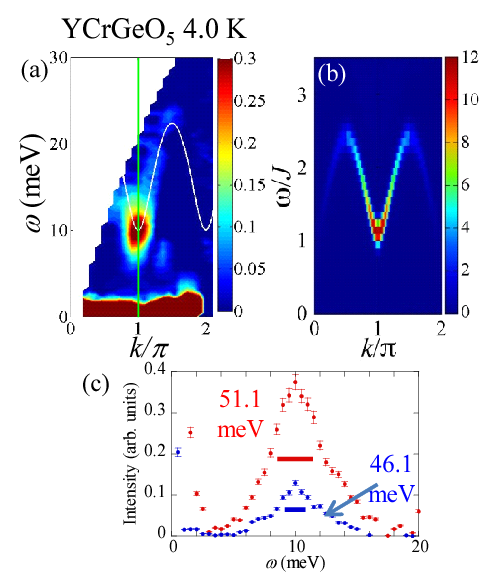}
\caption{
(Color online)
(a) Map of neutron scattering intensity in the $k - \omega$ plane 
of YCrGeO$_5$ at 4.0 K and $E_{\rm i} = 51.1$ meV 
obtained using the conversion method.\cite{Tomiyasu09} 
The horizontal axis indicates the normalized wavenumber parallel to the spin chain. 
The right vertical key shows the INS intensity in arbitrary units. 
The white line indicates 
$\omega (k) = \sqrt{A^2 \sin^2 k + \Delta^2}$ with 
$\Delta = 10$ meV and $A = 20$ meV.
(b) 
The dynamical structure factor of 
the AF Heisenberg alternating spin-3/2 chain with $\delta=0.75$ 
calculated using the dynamical DMRG method. 
The right vertical key shows the intensity in arbitrary units. 
We used a Lorentzian broadening 
with half width at half maximum $\eta=0.16J$.
(c) Experimental constant-$Q$ spectra at 
$k = \pi$ indicated by the green line in (a). 
Red and blue circles represent 
data at $E_{\rm i} = 51.1$ and 
46.1 meV (with a higher energy resolution), respectively.
The horizontal bars represent 
the energy resolution at $\omega = 0$ meV.
}
\end{center}
\end{figure}

\begin{figure}
\begin{center}
\includegraphics[width=8cm]{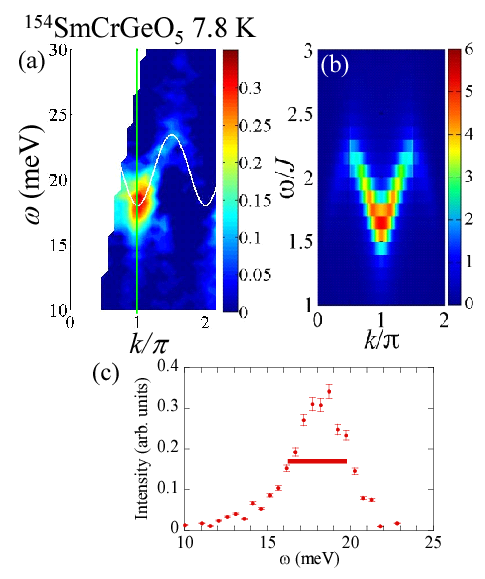}
\caption{
(Color online)
(a) Map of neutron scattering intensity in the $k - \omega$ plane 
of $^{154}$SmCrGeO$_5$ at 7.8 K and $E_{\rm i} = 91.6$ meV 
obtained using the conversion method.\cite{Tomiyasu09} 
The horizontal axis indicates the normalized wavenumber parallel to the spin chain.
The right vertical key shows the INS intensity in arbitrary units. 
The white line indicates 
$\omega (k) = \sqrt{A^2 \sin^2 k + \Delta^2}$ with 
$\Delta = 18$ meV and $A = 15$ meV.
(b) 
The dynamical structure factor of 
the AF Heisenberg alternating spin-3/2 chain with $\delta=0.9$ 
calculated using the dynamical DMRG method. 
The right vertical key shows the intensity in arbitrary units. 
We used a Lorentzian broadening 
with half width at half maximum $\eta=0.16J$.
(c) Experimental constant-$Q$ spectrum (circles) at 
$k = \pi$ indicated by the green line in (a). 
The horizontal bar represents
the energy resolution at $\omega = 0$ meV.
}
\end{center}
\end{figure}

We compare experimental INS intensities and 
calculated dynamical structure factors of 
the AF Heisenberg alternating spin-3/2 chains.  
The experimental values of $\omega (1.5 \pi) / \omega (\pi)$ 
are 2.2 and 1.3 in YCrGeO$_5$ and $^{154}$SmCrGeO$_5$, respectively.
When $\delta = 0.75$ and 0.90, 
values of $\omega (1.5 \pi) / \omega (\pi)$ are 2.2 and 1.3 
in the calculated results presented in Figs. 7(b) and 8(b), respectively. 
The calculated dynamical structure factors are similar to 
the experimental INS intensities. 
In the calculated result with $\delta = 0.75$, $\Delta / J = 1.1$. 
Therefore, $J = 9.1$ meV = 106 K in YCrGeO$_5$.
The ratio of the two exchange interaction values is 
$(1 - \delta)/(1 + \delta) = 0.14$.
In the calculated result with $\delta = 0.90$, $\Delta / J = 1.6$. 
Therefore, $J = 11$ meV = 128 K in $^{154}$SmCrGeO$_5$.
The value of  $(1 - \delta)/(1 + \delta)$ is 0.05.

Figures 7(c) and 8(c) show constant $Q$ spectra at $k = \pi$ of
YCrGeO$_5$ at 4.0 K
and $^{154}$SmCrGeO$_5$ at 7.8 K, respectively.
A single peak is apparent around the gap energy in each line. 
We were unable to estimate accurately the energy resolution around 10 meV and 18 meV 
because of powder samples. 
The energy resolution around 10 meV and 18 meV
is expected to be higher than that at 0 meV.
Therefore, we compared the width of the peak with 
the energy resolution at 0 meV (horizontal bar).
The width of the 10 meV peak in YCrGeO$_5$ 
is broader than the energy resolution. 
The 18 meV peak in $^{154}$SmCrGeO$_5$ 
is nearly resolution limited or might be slightly broader than the energy resolution. 
This result is consistent with the fact that 
the spin system of Cr$^{3+}$ spins 
in SmCrGeO$_5$ [$(1 - \delta)/(1 + \delta) = 0.05$]
is similar to an isolated AF dimer.

We show in Fig 3 that 
the experimental susceptibility is similar to the calculated one. 
The experimental $\chi_{\rm exp}$ of YCrGeO$_5$ consists of 
three terms $\chi_{\rm Cr}$, $\chi_{\rm CW}$, and $\chi_0$.
The first term $\chi_{\rm Cr}$ is 
susceptibility of the alternating spin chain with $J=106$ K and $\delta = 0.75$.
The second term $\chi_{\rm CW}$ is 
the Curie-Weiss term that is dominant at low $T$.
In the two terms, we reasonably assume that 
the $g$-value is 2 for Cr$^{3+}$ spins.  
From the data below 10 K, we obtained 
$\chi_{\rm CW} = \frac{0.052}{T-0.2}$ emu/mol, which 
means that about 2.8 \% \ of Cr spins are nearly isolated.
Therefore, the green line of $\chi_{\rm Cr}$ 
represents 0.972 of the molar susceptibility of the alternating spin chain.  
The third term $\chi_0$ is a constant term. 
When $\chi_0 = -1.3 \times 10^{-4}$ emu/mol, 
the sum of the three terms $\chi_{\rm cal}$ 
reproduces roughly the experimental $\chi_{\rm exp}$. 
We speculate that the negative value of $\chi_0$ 
originates mainly in non-magnetic Y$_2$Ge$_2$O$_7$. 
The sample used in the susceptibility measurement (2.8 \% \ isolated spins)
differs from 
that used in the high-field magnetization measurement
(5 \% \ isolated spins as aforementioned). 
We obtained a value close to 5 \% \ 
from susceptibility of a sample took from the same batch 
used in the high-field magnetization measurement.

The experimental $\chi_{\rm exp}$ of SmCrGeO$_5$ consists of 
three terms $\chi_{\rm Cr}$, $\chi_{\rm Sm}$, and $\chi_0$.
The first term $\chi_{\rm Cr}$ is 
susceptibility of the alternating spin chain with $J=128$ K and $\delta = 0.9$.
The second term $\chi_{\rm Sm}$ is 
the Curie-Weiss term generated by Sm$^{3+}$ ions.  
From the data below 30 K, we obtained 
$\chi_{\rm Sm} = \frac{0.055}{T+0.22}$ emu/mol.
The ground state of Sm$^{3+}$ ions is $^6 H_{5/2}$, meaning that 
the value of ${\bm J}$ ($= {\bm L} + {\bm S}$) is $|L-S| = 5/2$.  
Here, ${\bm L}$ is the total angular momentum.
The value of the Land\'e $g$ factor is 2/7.
Thus,  the Curie constant of Sm$^{3+}$ ions
is calculated as 0.0893 emu K/mol, 
which is slightly larger than the experimental value.  
The third term $\chi_0$ is a constant term. 
When $\chi_0 = 1.6 \times 10^{-3}$ emu/mol, 
the sum of the three terms $\chi_{\rm cal}$
reproduces roughly the experimental $\chi_{\rm exp}$. 

Our INS and susceptibility results indicate that 
the dominant $J(1+ \delta ) (\equiv J_1)$ interaction is AF. 
In SmCrGeO$_5$, the other $J(1- \delta ) (\equiv J_2)$ interaction (12.8 K)
is much smaller than the $J_1$ interaction (243 K), 
indicating that 
the spin system is regarded as weakly-coupled AF dimers. 
Therefore, even if the $J_2$ interaction is ferromagnetic (F), 
our conclusion is intrinsically unchanged. 
In YCrGeO$_5$, we infer that the $J_2$ interaction must be AF. 
As described, 
the alternating spin chain with $J_1 = 186$ K and $J_2 = 26.5$ K 
can explain our INS and susceptibility results.
In addition, the Cr-O-Cr angles suggest that 
the $J_2$ interaction in YCrGeO$_5$ is AF as follows. 
An exchange interaction value is determined mainly by the Cr-O-Cr angle. 
The angle in the $d_1$ bond is larger than that in the $d_2$ bond. 
Therefore, the $J_1$ interaction exists in the $d_1$ bond. 
The angles in the $d_1$ bond are 95.9$^{\circ}$ and 97.3$^{\circ}$ for 
YCrGeO$_5$ ($J_1 = 186$ K) and SmCrGeO$_5$ ($J_1 = 243$ K), respectively. 
The smaller angle is consistent with the smaller $J_1$ value. 
The angle in the $d_2$ bond is larger in YCrGeO$_5$ (92.8$^{\circ}$) than 
SmCrGeO$_5$ (91.3$^{\circ}$). 
If the $J_2$ interaction in YCrGeO$_5$ is F, 
then the $J_2$ interaction in SmCrGeO$_5$ is also F and 
the magnitude $|J_2|$ must be larger 
in SmCrGeO$_5$ than in YCrGeO$_5$, 
which is inconsistent with the INS results (stronger dispersion in YCrGeO$_5$). 
Consequently, it is natural to infer that 
the $J_2$ interaction in YCrGeO$_5$ is AF.

\section{Conclusion}

We conducted inelastic neutron scattering (INS) and 
magnetization measurements 
on {\it R}CrGeO$_5$ ({\it R} = Y or Sm) powders. 
The high-field magnetization of YCrGeO$_5$ suggests 
the existence of a spin gap.
We observed spin gap (singlet-triplet) excitations and  
the dispersion relation of the lowest magnetic excitations in the INS results.  
The experimental results are consistent with 
the calculated results of 
the antiferromagnetic alternating spin-3/2 chain. 
YCrGeO$_5$ and SmCrGeO$_5$ are 
the first spin-3/2 chain substances 
having a spin-singlet ground state with a spin gap.
From the alternation ratio, 
the ground state is expected to be the $(2S, 0)$ state as shown in Fig. 1. 
We will study other {\it R}CrGeO$_5$. 
We expect to find substances of which 
ground state is the $(2S-1, 1)$ state as shown in Fig. 1. 

\begin{acknowledgments}

The neutron scattering experiments on $^{154}$SmCrGeO$_5$ 
were approved by 
the Neutron Scattering Program Advisory Committee of IMSS, KEK 
(Proposal No. 2013S01).
The neutron scattering experiments on YCrGeO$_5$ 
were approved by 
the Neutron Science Proposal Review Committee of J-PARC/MLF 
(Proposal No. 2012B0009) and 
supported by the Inter-University Research Program on 
Neutron Scattering of IMSS, KEK.
The high-field magnetization experiments were conducted under 
the Visiting Researcher's Program of 
the Institute for Solid State Physics, 
the University of Tokyo.
This work was supported by grants from NIMS. 
We are grateful 
to S. Matsumoto for sample syntheses and X-ray diffraction measurements,  
to M. Kaise for X-ray diffraction measurements, and 
to H. Sakurai for fruitful discussion.  
M.K. thanks S. Yamada and S. Nishimoto 
for helpful discussions on numerical techniques of the DMRG method. 
The theoretical work was supported by KAKENHI (No. 23540428) and 
the World Premier International Research Center Initiative (WPI), MEXT, Japan. 

\end{acknowledgments}

\newpage %Just because of unusual number of tables stacked at end

\begin{references}

\bibitem{Kato94}
Y. Kato and A. Tanaka,
J. Phys. Soc. Jpn. {\bf 63}, 1277 (1994).

\bibitem{Kohno98}
M. Kohno, M. Takahashi, and M. Hagiwara, 
Phys. Rev. B {\bf 57}, 1046 (1998).

\bibitem{Yajima96}
M. Yajima and M. Takahashi, 
J. Phys. Soc. Jpn. {\bf 65}, 39 (1996).

\bibitem{Yamamoto97} 
S. Yamamoto, 
Phys. Rev. B {\bf 55}, 3603 (1997).

\bibitem{Diederix79}
K. M. Diederix, H. W. J. Bl\"ote, J. P. Groen, T. O. Klaassen, and N. J. Poulis, 
Phys. Rev. B {\bf 19}, 420 (1979).

\bibitem{Bonner83}
J. C. Bonner, S. A. Friedberg, H. Kobayashi, D. L. Meier, and H. W. J. Bl\"ote, 
Phys. Rev. B {\bf 27}, 248 (1983).

\bibitem{Bray75}
J. W. Bray, H. R. Hart, Jr., L. V. Interrante, I. S. Jacobs, J. S. Kasper, G. D. Watkins, S. H. Wee, and J. C. Bonner, 
Phys. Rev. Let. {\bf 35}, 744 (1975).

\bibitem{Jacobs76}
I. S. Jacobs, J. W. Bray, H. R. Hart, Jr., L. V. Interrante, J. S. Kasper, and G. D. Watkins, D. E. Prober, and J. C. Bonner, 
Phys. Rev. B {\bf 14}, 3036 (1976).

\bibitem{Hase93a}
M. Hase, I. Terasaki, and K. Uchinokura, 
Phys. Rev. Lett. {\bf 70}, 3651 (1993).

\bibitem{Hase93b}
M. Hase, I. Terasaki, Y. Sasago, K. Uchinokura, and H. Obara, 
Phys. Rev. Lett. {\bf 71}, 4059 (1993).

\bibitem{Hase93c}
M. Hase, I. Terasaki, K. Uchinokura, M. Tokunaga, 
N. Miura, and H. Obara, 
Phys. Rev. B {\bf 48}, 9616 (1993).

\bibitem{Narumi01}
Y. Narumi, M. Hagiwara, M. Kohno, and K. Kindo, 
Phys. Rev. Let. {\bf 86}, 324 (2001).

\bibitem{Zheludev04}
A. Zheludev, T. Masuda, B. Sales, D. Mandrus, T. Papenbrock, T. Barnes, and S. Park, 
Phys. Rev. B {\bf 69}, 144417 (2004).

\bibitem{Renard87}
J. P. Renard, M. Verdaguer, L. P. Regnault, W. A. C. Erkelens, J. Rossat-Mignod, and W. G. Stirling, 
Europhys. Lett. {\bf 3}, 945 (1987).

\bibitem{Darriet93}
J. Darriet and L. P. Regnault, 
Solid State Commun. {\bf 86}, 409 (1993).

\bibitem{DiTusa94}
J. F. DiTusa, S-W. Cheong, C. Broholm, G. Aeppli, L. W. Rupp, Jr., and B. Batlogg, 
Physica B {\bf 194-196}, 181 (1994).

\bibitem{Yokoo95}
T. Yokoo, T. Sakaguchi, K. Kakurai, and J. Akimitsu, 
J. Phys. Soc. Jpn. {\bf 64}, 3651 (1995).

\bibitem{Granroth96}
G. E. Granroth, M. W. Meisel, M. Chaparala, Th. Jolicoeur, B. H. Ward, and 
D. R. Talham, 
Phys. Rev. Let. {\bf 77}, 1616 (1996).

\bibitem{Hagiwara12}
M. Hagiwara, Y. Idutsu, Z. Honda, and S. Yamamoto, 
J. Phys.: Conf. Ser. {\bf 400}, 032014 (2012).

\bibitem{Granroth02}
G. E. Granroth, S. E. Nagler, R. Coldea, R. S. Eccleston, B. H. Ward, 
D. R. Talham, and M. W. Meisel, 
App. Phys. A {\bf 74}, S868 (2002).

\bibitem{Kakurai91}
K. Kakurai, M. Steiner, R. Pynn, and J. K. Kjems,
J. Phys.: Condens. Matter {\bf 3}, 715 (1991).

\bibitem{Itoh12a}
S. Itoh, T. Yokoo, S. Yano, D. Kawana, H. Tanaka, and Y. Endoh, 
J. Phys. Soc. Jpn. {\bf 81}, 084706 (2012).

\bibitem{Itoh99}
S. Itoh, Y. Endoh, K. Kakurai, H. Tanaka, S. M. Bennington, T. G. Perring, 
K. Ohoyama, M. J. Harris, K. Nakajima, and C. D. Frost, 
Phys. Rev. B {\bf 59}, 14406 (1999).

\bibitem{Itoh02}
S. Itoh, H. Tanaka, and M. J. Bull, 
J. Phys. Soc. Jpn. {\bf 71}, 1148 (2002).

\bibitem{Shpanchenko08}
R. V. Shpanchenko, A. A. Tsirlin, E. S. Kondakova, E. V. Antipov, C. Bougerol, J. Hadermann, G. van Tendeloo, H. Sakurai, and E. Takayama-Muromachi, 
J. Solid State Chem. {\bf 181}, 2433 (2008).

\bibitem{Itoh11}
S. Itoh, T. Yokoo, S. Satoh, S. Yano, D. Kawana, J. Suzuki, and T. J. Sato, 
Nucl. Instr. Meth. Phys. Research A {\bf 631}, 90 (2011).

\bibitem{Yano11}
S. Yano, S. Itoh, T.Yokoo, S. Satoh, T. Yokoo, D. Kawana, and T. J. Sato,
Nucl. Instr. Meth. Phys. Research A {\bf 654}, 421 (2011).

\bibitem{Itoh12b}
S. Itoh, K. Ueno, and T. Yokoo, 
Nucl. Instr. Meth. Phys. Research A {\bf 661}, 58 (2012).

\bibitem{Evertz03}
H. G. Evertz, Adv. Phys. {\bf 52}, 1 (2003).

\bibitem{Jeckelmann02}
E. Jeckelmann, Phys. Rev. B {\bf 66} 045114 (2002).

\bibitem{Tomiyasu09}
K. Tomiyasu, M. Fujita, A. I. Kolesnikov, R. I. Bewley, M. J. Bull, and S. M. Bennington, 
Appl. Phys. Lett. {\bf 94}, 092502 (2009).

\end{references}

\end{document}